\documentclass[11pt]{article}

\usepackage{amssymb}
\usepackage{graphicx}
\usepackage[numbers,compress]{natbib}
\begin{document}
	
	\title{Masses dynamics during black hole evaporation in spatially flat FLRW universes}
	
	\author{Ion I. Cot\u aescu\thanks{Corresponding author E-mail:~~i.cotaescu@e-uvt.ro} and Ion Cot\u aescu Jr,\\
		{\it West University of Timi\c soara,} \\{\it V. Parvan Ave. 4,
			RO-300223 Timi\c soara}}

\maketitle

\begin{abstract}
The new dynamical solutions of  Einstein's equations with perfect fluid in space-times with FLRW asymptotic behaviour,  derived recently  \cite{Cot,Cot1,Cot2},  may describe evaporating black holes  whose masses dissipate into dust. The time evolution of the black hole and dust masses of simple models of evaporating black holes are studied  using analytical and graphical methods.

Pacs: 04.70.Bw
\end{abstract}


\section{Introduction}

The  Friedmann-Lema\^ itre-Robertson-Walker (FLRW) space-times are the principal models of universe considered in actual cosmology. These universes  may be populated  either with static black holes, which are vacuum solutions of Einstein's equations \cite{BH}, or with dynamical particles defined as isotropic solutions of Einstein's equations with perfect fluid, laying out a FLRW asymptotic behaviour (see for instance Ref. \cite{TB1}).  The first metric of dynamical particles behaving as black holes was proposed by McVittie long time ago \cite{MV} and then studied by many authors \cite{MV1,MV2,MV3,MV4,A,A1,A2}.  These geometries lay out curved space sections being produced by the pressure of the perfect fluid which is singular on the Schwarzschild sphere while its density remains just that of the asymptotic FLRW space-time. Other models of  inhomogeneous dynamical spherically symmetric solutions of Einstein equations were studied  so far \cite{A,A1,A2,A3}. Note that the charged version of the McVittie metric was proposed by Shah and Vaidya  \cite{V1,V2} and then considered by other authors \cite{V3,V4,V5,V6,V7}.

In this stream, new models of dynamical neutral black holes were proposed recently describing systems constituted by a Schwarzschild-type black holes surrounded by a cloud of dust, hosted by the perfect fluid of the background which determines a  FLRW asymptotic behaviour. When the space-time is expanding then  the black hole evaporates dissipating its mass into dust but without affecting the FLRW fluid.  Remarkably, in these models the Hubble function of the FLRW asymptotic space-time is proportional through a real valued constant, denoted by $\kappa$, to the total mass of the dynamical system black hole-dust.  Based on this  property the black hole evaporation in de Sitter expanding universe was studied  showing that the entire black hole mass dissipates into dust but conserving the total black hole-dust mass \cite{Cot2}. This model shows how  these black holes evaporate naturally because of their geometries without involving supplemental mechanism as, for example,  the Hawking radiation.  

In the present paper we would like to extend this study to other simple models of dynamical systems black hole-dust whose total mass is no longer conserved, being adsorbed by the FLRW fluid.  We focus on the models having simple FLRW scale factors proportional with powers of the cosmic time. Such models were briefly inspected in Ref. \cite{Cot1} where only the profiles of the black hole and cosmological horizons were studied but without analyzing the mass dynamics. Here we show how the black-hole and dust masses of these models evolve in time proving that finally the evaporation is complete and the entire dust mass is adsorbed by the FLRW fluid.

We start in the next section presenting the new solutions of Einstein's equations we call $\kappa$-models pointing out the gravitational sources, the density and pressure of the whole fluid, and giving the equation of state relating the Hubble function of the asymptotic FLRW space-time to the total mass of the system black hole-dust. In the next section we focus on the aforementioned simple models studying the masses dynamics using analytical and graphical methods. Moreover, we observe  that the black hole and dust masses are not accessible to an observer staying in the physical domain  which can measure only the evolution of the local dust density. We show that in the physical domain the profile of this quantity feels the black hole singularity and depend on the parameter $\kappa$. Finally we present our conclusions.

 We use the Planck units with $\hbar=c=G=1$.

\section{Dynamical particles in physical frames}

For studying  static or dynamic non-rotating black holes  in space-times with spherical symmetry and spatially flat sections,  it is convenient to use the physical frames $\{t, {\bf x}\}$  with  Painlev\' e-Gullstrand  coordinates  \cite{Pan,Gul},  $x^{\mu}$ ($\alpha,\mu,\nu,...=0,1,2,3$) formed by the {\em cosmic time}, $x^0=t$, and physical Cartesian space coordinates,  ${\bf x}=(x^1,x^2,x^3)$, associated to the spherical ones $(r,\theta,\phi)$. The models we intend to study here have  line elements  in physical frames of the general form  
\begin{eqnarray}
	ds^2&=& g_{\mu\nu}(x)dx^{\mu}dx^{\nu}=dt^2 -\left[dr-h (t,r)dt\right]^2-r^2 d\Omega^2 \nonumber\\
	&=&\left[1-h (t,r)^2\right]dt^2+2 h(t,r)dr dt -dr^2-r^2 d\Omega^2\,,\label{fam}
\end{eqnarray} 
depending on the smooth functions $h$ and $d\Omega^2=d\theta^2+\sin^2\theta\, d\phi^2$. 

In what follows we focus on the $\kappa$-models whose $h$-functions in proper co-moving frames where the four-velocity has the components 
\begin{eqnarray}
	U_{\mu}&=&\left(\frac{1}{\sqrt{g^{00}}},0,0,0\right)=(1,0,0,0)\,, \\ 
	U^{\mu}&=&g^{\mu 0}U_0=\frac{g^{\mu 0}}{\sqrt{g^{00}}} ~~\Rightarrow~~U^{\mu}=(1,h,0,0)\,,
\end{eqnarray}
have the form \cite{Cot1}
\begin{equation}\label{hdef}
	h_{\kappa}(t,r)=-\frac{1}{3}\frac{\dot{M}(t)}{M(t)} r+\epsilon\sqrt{\frac{2 M(t)}{r}+\kappa^2 M(t)^2 r^2}\,,
\end{equation}
depending on the dynamical mass, $M(t)$, its time derivative $\dot M(t)=\partial_t M(t)$,  and the  constant $\kappa=\epsilon|\kappa|$  ($\epsilon={\rm sign}\, \kappa$) playing the role of free parameter.  

Denoting the space-times of these models as ${\frak M}_{\epsilon}(M,\kappa)$, we find that these have isotropic Einstein tensors whose components in the proper co-moving physical frames have the form \cite{Cot1}
\begin{eqnarray}
	G^0_0(M,\kappa)&=&3\kappa^2 M(t)^2 +\frac{1}{3} \frac{\dot M(t)^2}{M(t)^2}\nonumber\\
	&-&2\epsilon\dot{M}(t)\frac{1+{M(t)}\kappa^2r^{3} }{\sqrt{M(t)(M(t)\kappa^2r^3+2) r^3}}=8\pi \rho_{\kappa}\,,\label{E1}\\
G(M,\kappa)&\equiv&G^r_r(M,\kappa)=G^{\theta}_{\theta}(M,\kappa)=G^{\phi}_{\phi}(M,\kappa)\nonumber\\
&=&3\kappa^2 M(t)^2+ \frac{\dot M(t)^2}{M(t)^2}-\frac{2}{3} \frac{\ddot M(t)}{M(t)}=-8\pi p_{\kappa}	\,,\label{E2}
\end{eqnarray}
satisfying, in addition, the condition
\begin{equation}
G^0_r(M,\kappa)=0 \Rightarrow G^r_0(M,\kappa)=\frac{g^{0r}}{g^{00}}\left(G^0_0(M,\kappa)-G(M,\kappa)\right)\,.\label{cond}
\end{equation}
The conclusion is that the $\kappa$-models are solutions of Eistein's equations with perfect fluid having the density  $\rho_{\kappa}$ and pressure $p_{\kappa}$.

The asymptotic space-time of  ${\frak M}(M,\kappa)$, for $r\to\infty$,  is a FLRW manifold  with a scale factor $a(t)$, denoted by ${\frak M}(a)$, whose Hubble function is defined by the asymptotic condition 
\begin{equation}\label{hubb}
\frac{\dot a(t)}{a(t)}=\lim_{r\to\infty}\frac{h_{\kappa}(t,r)}{r}=\kappa M(t) -\frac{1}{3}\frac{\dot{M}(t)}{M(t)}\,.
\end{equation}
The metric tensors of  ${\frak M}(a)$, give the asymptotic line elements in physical frames of the form,
\begin{equation}\label{s2}
	ds^2=\left(1-\frac{\dot a^2}{a^2}\, r^2\right)dt^2+2\frac{\dot a}{a}\, r\, dr\, dt -dr^2-r^2d\Omega^2\,,
\end{equation}
 emphasizing the apparent horizon on the sphere of radius  
 \begin{equation}\label{ra}
 	r_a(t)=\left| \frac{a(t)}{\dot a(t)}\right| \,,
 \end{equation}
 we call the asymptotic horizon. 
 Integrating Eq. (\ref{hubb})  with the initial condition $a(t_0) =1$ we obtain the expression of the scale factor  \cite{Cot1}
 \begin{equation}\label{at1}
 	a(t)=\left(  \frac{M_0}{M(t)}\right)^{\frac{1}{3}} \exp\left( \kappa\int_{t_0}^t M(t')dt'\right)
 \end{equation}
 of the asymptotic FLRW space-time  where we denoted $M_0=M(t_0)$.
Reversely, we can start with the scale factor deriving then the mass function as
 \begin{equation}\label{mat} 
 	M(t)=\frac{M_0}{a(t)^3}\left[ 1-3\kappa M_0 \int_{t_0}^t \frac{dt'}{a(t')^3}\right]^{-1}\,.
 \end{equation} 
Using this formula we have to avoid the singularities  imposing the restriction  $\kappa\in[0,k_{lim}]$ where $k_{lim}$ must satisfy \cite{Cot1}
 \begin{equation}\label{coco}
 	3\kappa_{lim} M_0 \int_{t_0}^\infty \frac{dt'}{a(t')^3}	=1\,,
 \end{equation}  
 because the function $a(t)$ is positively defined.
 
 The Hubble functions must depend monotonously of time, without zeros in the physical time domain which  might produce singularities of the function $r_a(t)$ giving the radius of the asymptotic horizon (\ref{ra}).  We may prevent  these zeros to appear imposing different conditions  for expanding or collapsing  space-times as \cite{Cot1}
 \begin{eqnarray}
 	{\rm expanding:} &~~~~&\frac{\dot a(t)}{a(t)}>0 ~~\Rightarrow ~~ 	\frac{\dot{M}(t)}{M(t)}<0\,,\quad \epsilon=1\,,\label{expand}\\
 	{\rm collapsing:} &~~~~&	\frac{\dot a(t)}{a(t)}<0 ~~\Rightarrow ~~ 	\frac{\dot{M}(t)}{M(t)}>0\,,\quad \epsilon=-1\,.\label{collaps}
 \end{eqnarray} 
 Note that the expanding space-times are of interest in cosmology. 
 
 The physical meaning of the $\kappa$-models results from the right-handed sides of Einstein's equations (\ref{E1}) and (\ref{E2}). First of all, we observe that there are no static terms which means that we do not need to consider a cosmological constant. Furthermore, we separate the density and pressures of the FLRW fluid 
 \begin{eqnarray}
 	\rho_a(t)&=&\frac{3}{8\pi}\left(\frac{\dot a}{a}\right)^2  =\frac{1}{8\pi}\left( 3\kappa^2 M(t)^2 +\frac{1}{3} \frac{\dot M(t)^2}{M(t)^2}-2\kappa \dot M(t)\right)\,,\\
 	p_a(t)&=&- \frac{1}{8\pi}\left[ 3\left(\frac{\dot a}{a}\right)^2 +2\frac{d}{dt}\left(\frac{\dot a}{a}\right)   \right] \nonumber\\
 	& =&-\frac{1}{8\pi}\left( 3\kappa^2 M(t)^2+ \frac{\dot M(t)^2}{M(t)^2}-\frac{2}{3} \frac{\ddot M(t)}{M(t)} \right)\,,
 \end{eqnarray}
 which depend only on the Hubble function. Hereby it results that the total density $\rho_{\kappa}=\rho_a + \delta\rho$  of the perfect fluid of the space-time ${\frak M}(M,\kappa)$ gets the new point-dependent  term \cite{Cot1}
 \begin{equation}
 	\delta\rho(t,r)=\frac{1}{4\pi}\epsilon \dot M(t)\left[ | \kappa|  -\frac{1+{M(t)}\kappa^2r^{3} }{\sqrt{M(t)(M(t)\kappa^2r^3+2) r^3}}   \right] \,, \label{dust}	
 \end{equation}
 while its pressure remains unchanged, $p_{\kappa}=p_a$. This means that $\delta\rho$ is the density of an amount of {\em dust} which does not modify the pressure $p_a$ of the FLRW fluid.

We have thus the image of a system formed by a dynamical black hole of Schwarzschild type, surrounded by a cloud of dust hosted by the homogeneous perfect fluid of a FLRW space-time. It is remarkable that the features of all these components are determined only by the function $M(t)$ and the parameter $\kappa$. The black hole is produced by the typical  singular term  $\frac{2M(t)}{r}$ of the function (\ref{hdef}) while the dust density (\ref{dust}) was obtained solving the Einstein equations. Observing that this density also is singular in origin, behaving as $\sim r^{-\frac{2}{3}}$, we may ask which is the relation or  interaction between the black hole and dust predicted by our model.
For solving this problem we derived the total mass of dust \cite{Cot2},
\begin{eqnarray}
	\delta M(t)	=4\pi \int_{0}^{\infty} dr\,r^2 \delta\rho(t,r) =
	-\frac{1}{3\kappa}\frac{\dot M(t)}{M(t)}\,,\label{dM}
\end{eqnarray}  
which is finite for $\kappa\not=0$. This  result allows us to rewrite Eq. (\ref{hubb}) as
\begin{equation}\label{aMM}
	\frac{\dot a(t)}{a(t)}=\kappa\left[	M(t) +\delta M(t) \right]\equiv \kappa m(t)\,,
\end{equation}	
relating thus the principal quantities which determine the behaviour of our model and introducing the total mass $m(t)$ of the system black hole-dust. This relation can be seen as a conservation law or even as an equation of state we may use instead of the traditional one ($p=w\rho$) which does not make sense in the case of our $\kappa$-models where $\rho_{\kappa}$ is point-dependent while $p_{\kappa}$ is homogeneous.  

 { \begin{figure}
		\centering
		\includegraphics[scale=0.21]{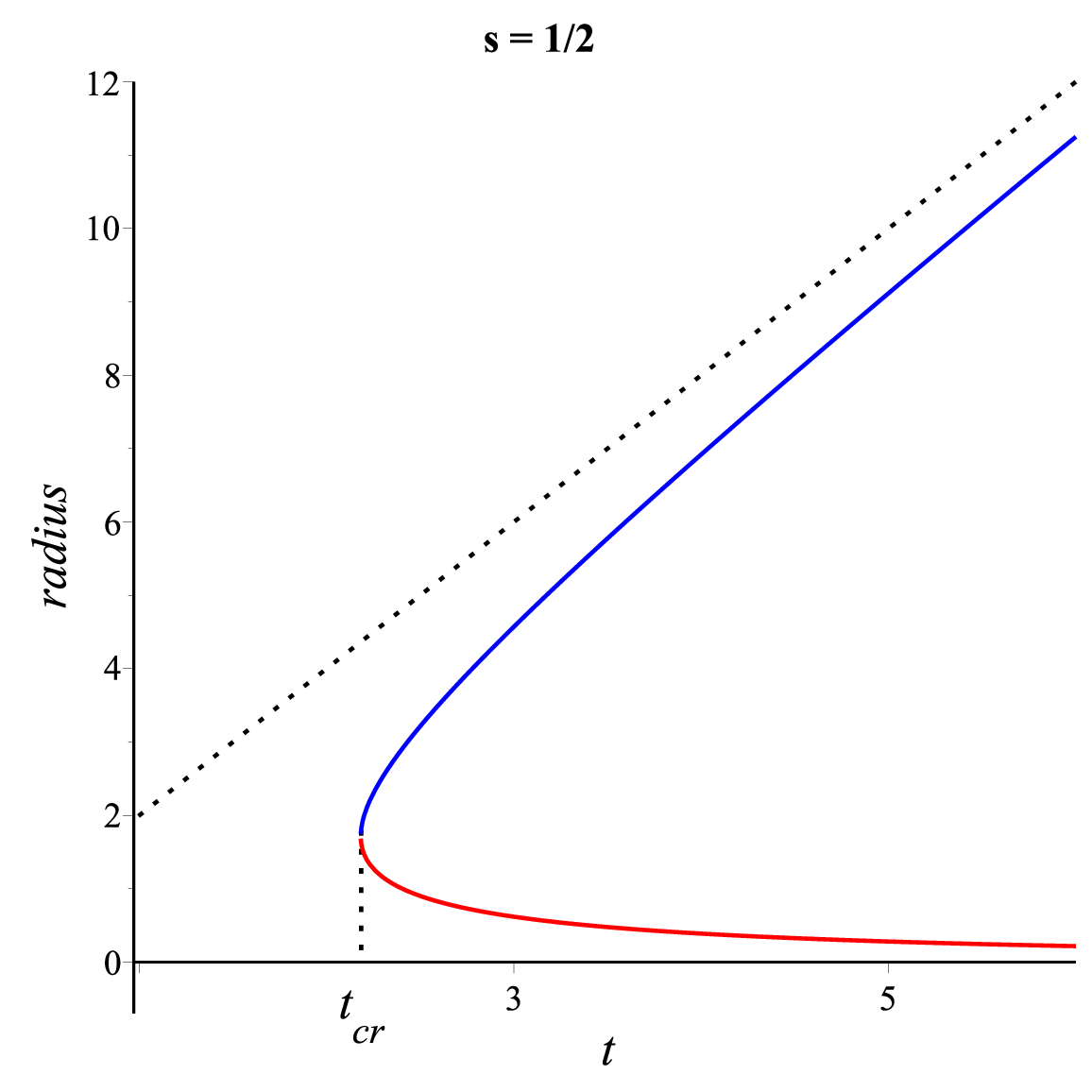}
		\includegraphics[scale=0.21]{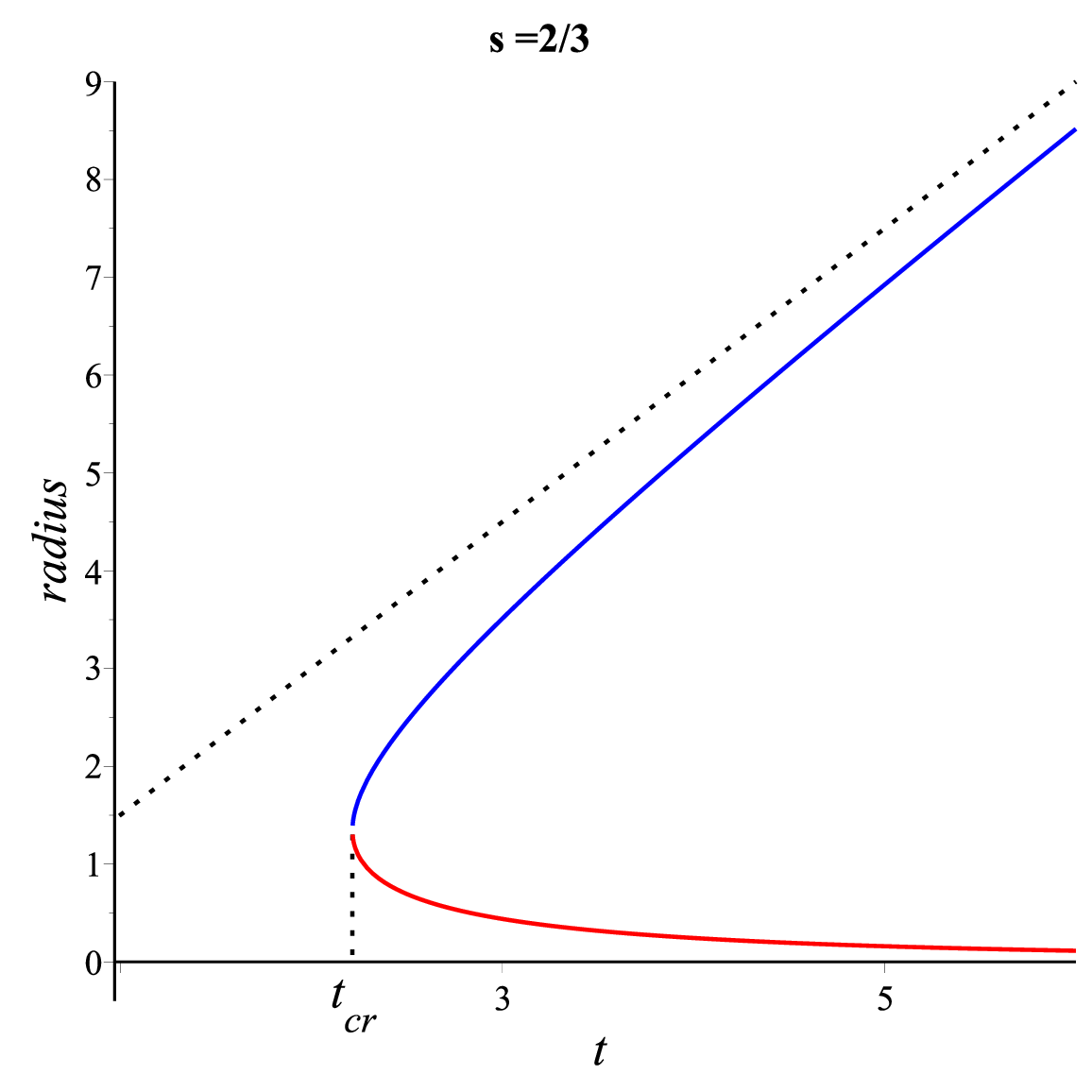}	
		\includegraphics[scale=0.21]{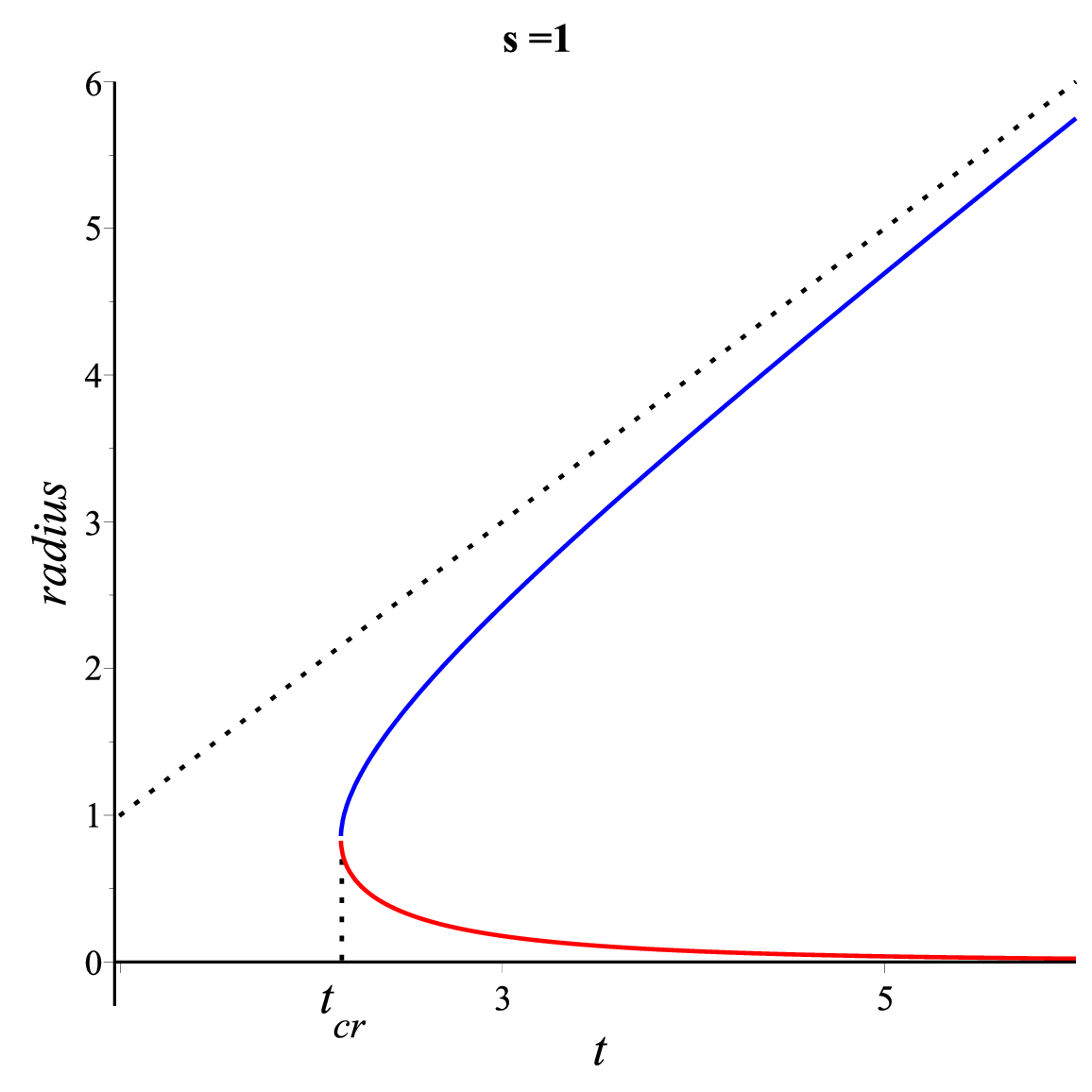}	
		\caption{The time evolution of the black hole (lower line) and cosmological (uper line) horizons of the models under consideration forming C-curves. The dotted lines represent the asymptotic horizons which are linear.}
\end{figure}}  

\section{Mass evolution in simple $\kappa$-models}

Let us see now how the black hole and dust masses evolve in simple $\kappa$-models whose asymptotic FLRW space-times have scale factors of the form
\begin{equation}\label{aaa}
	a(t)=\left(\frac {t}{t_0}\right)^s\,, \quad s>0\,,
\end{equation} 
complying with the initial condition $a(t_0)=1$. Here the time $t=0$ seems to play the role of big bang time but it is possible that this instant should not have physical meaning. This may happens because the cosmic time is a genuine  time-like variable only in the physical domain where $g_{00}(t,r)>0$.    This domain is bordered by the black hole and cosmological horizons whose radii, $r_b(t)$ and respectively $r_c(t)$,  are solutions of the equation $g_{00}(t,r)=0$. We have seen that these horizons arise on the same sphere at a critical time $t_{cr}>0$ when $r_b(t_{cr})=r_c(t_{cr})$ \cite{Cot,Cot1}. For $t>t_{cr}$  the black hole horizon collapses to zero while the cosmological one is expanding tending to the asymptotic horizon whose radius $r_a(t)$ is just the inverse of the Hubble function. In the case of the models having the scale factors (\ref{aaa}) this is linear in time, $r_a(t)=s^{-1}t$, while the black hole and cosmological horizons form C-curves  defining the physical domain, $t \ge t_{cr}, \, r_b(t)<r<r_c(t)$. As the horizons of similar models  were studied in Refs. \cite{Cot,Cot1}, here we restrict ourselves to present only the examples plotted in Fig.1.  

{ \begin{figure}
		\centering
		\includegraphics[scale=0.21]{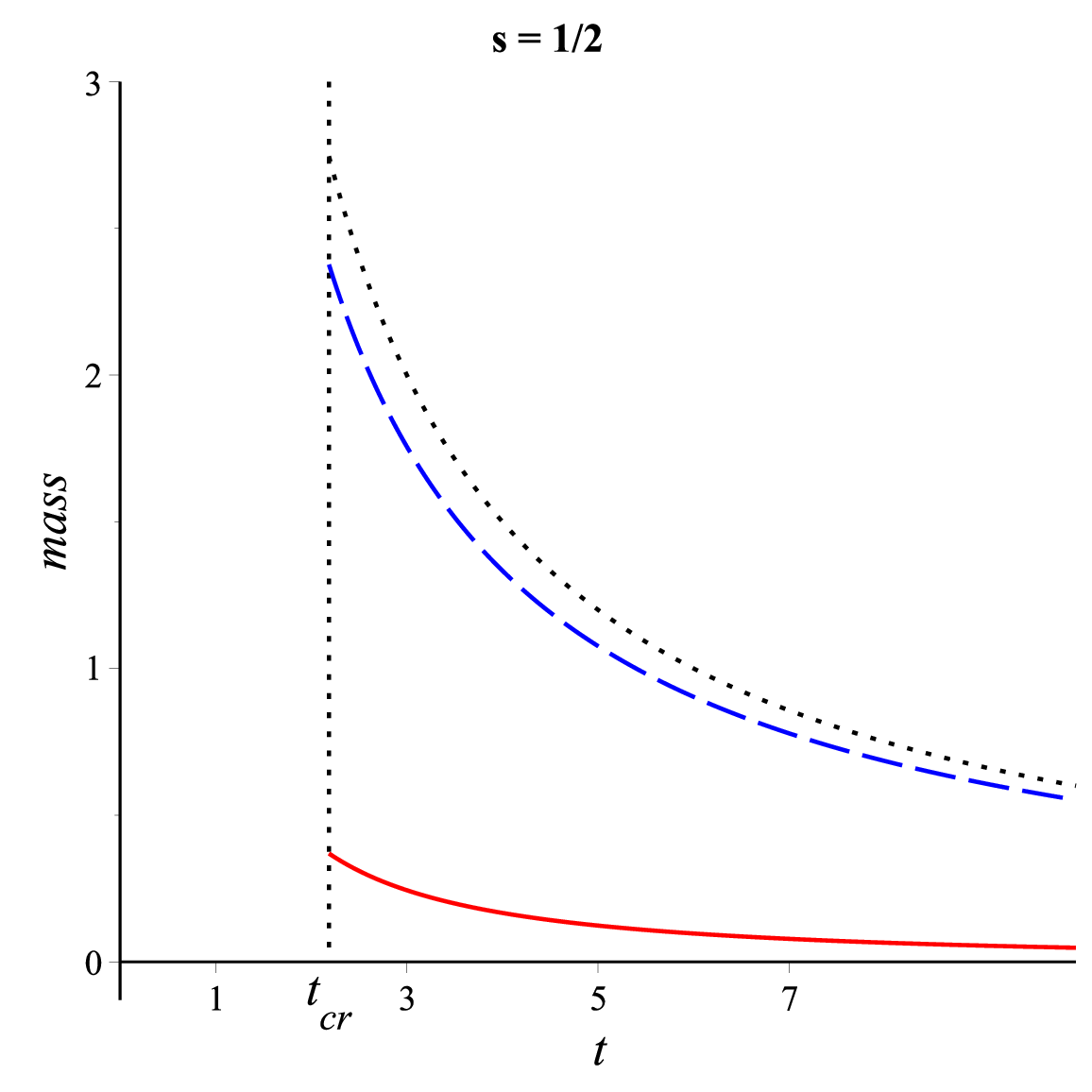}
		\includegraphics[scale=0.21]{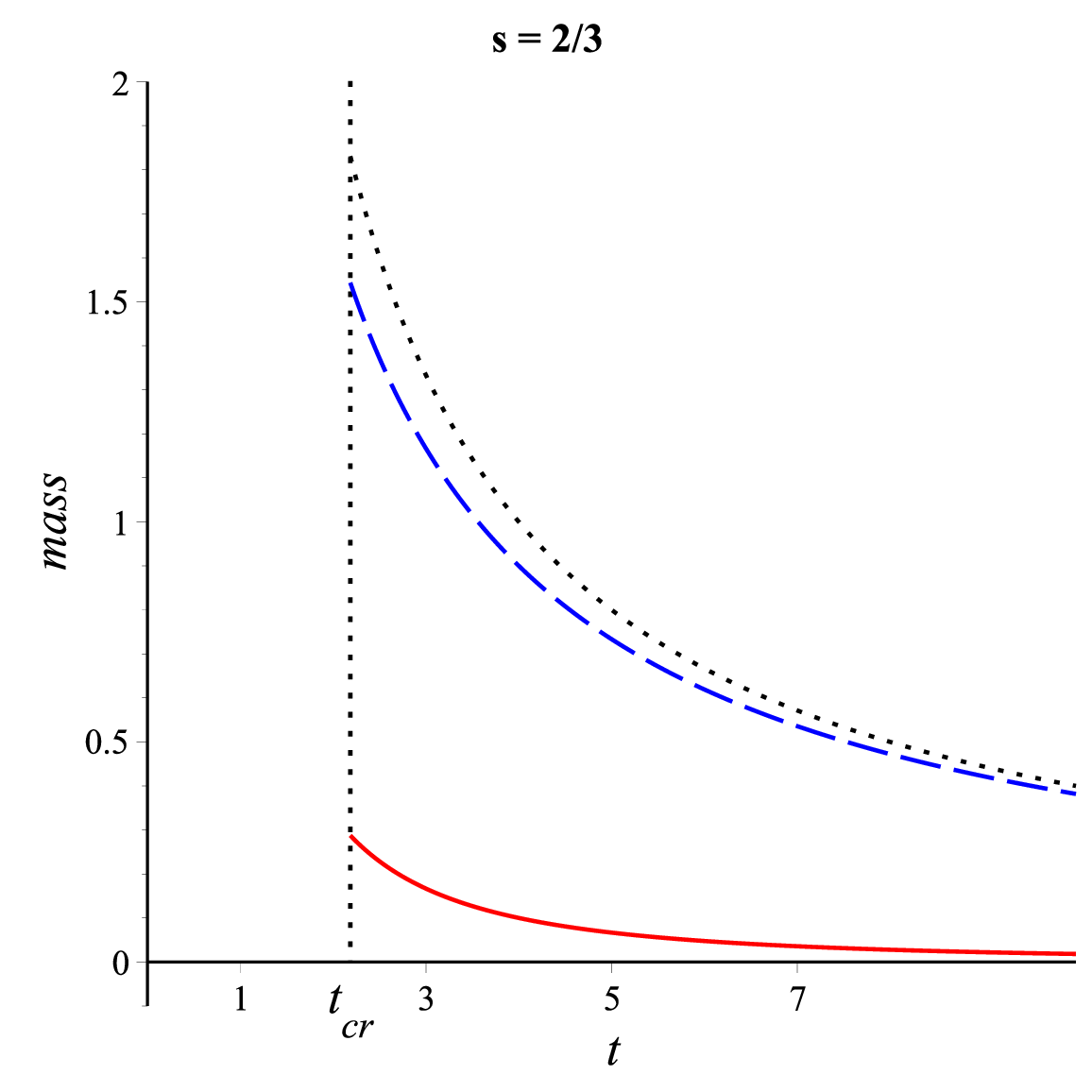}	
		\includegraphics[scale=0.21]{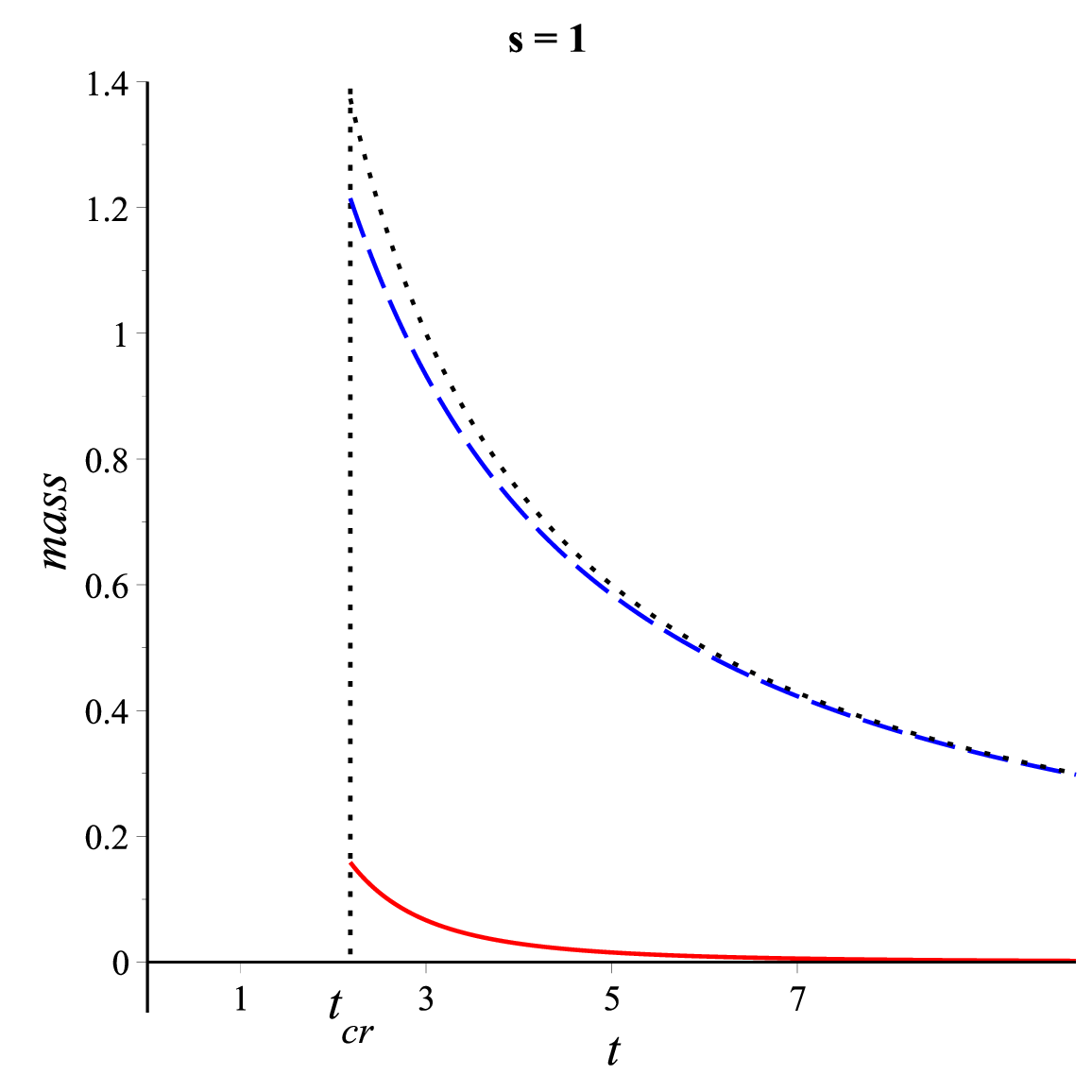}
		\includegraphics[scale=0.21]{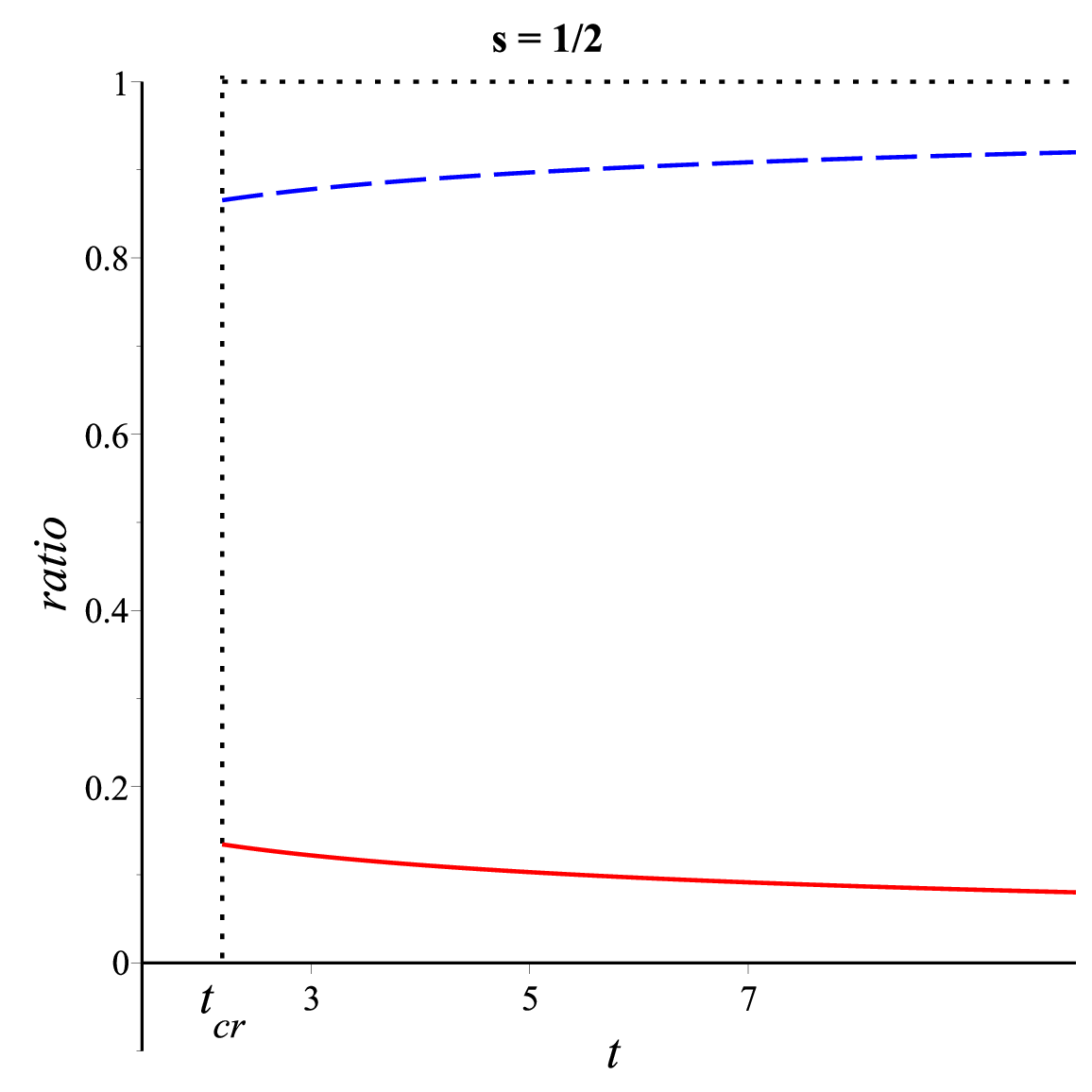}
		\includegraphics[scale=0.21]{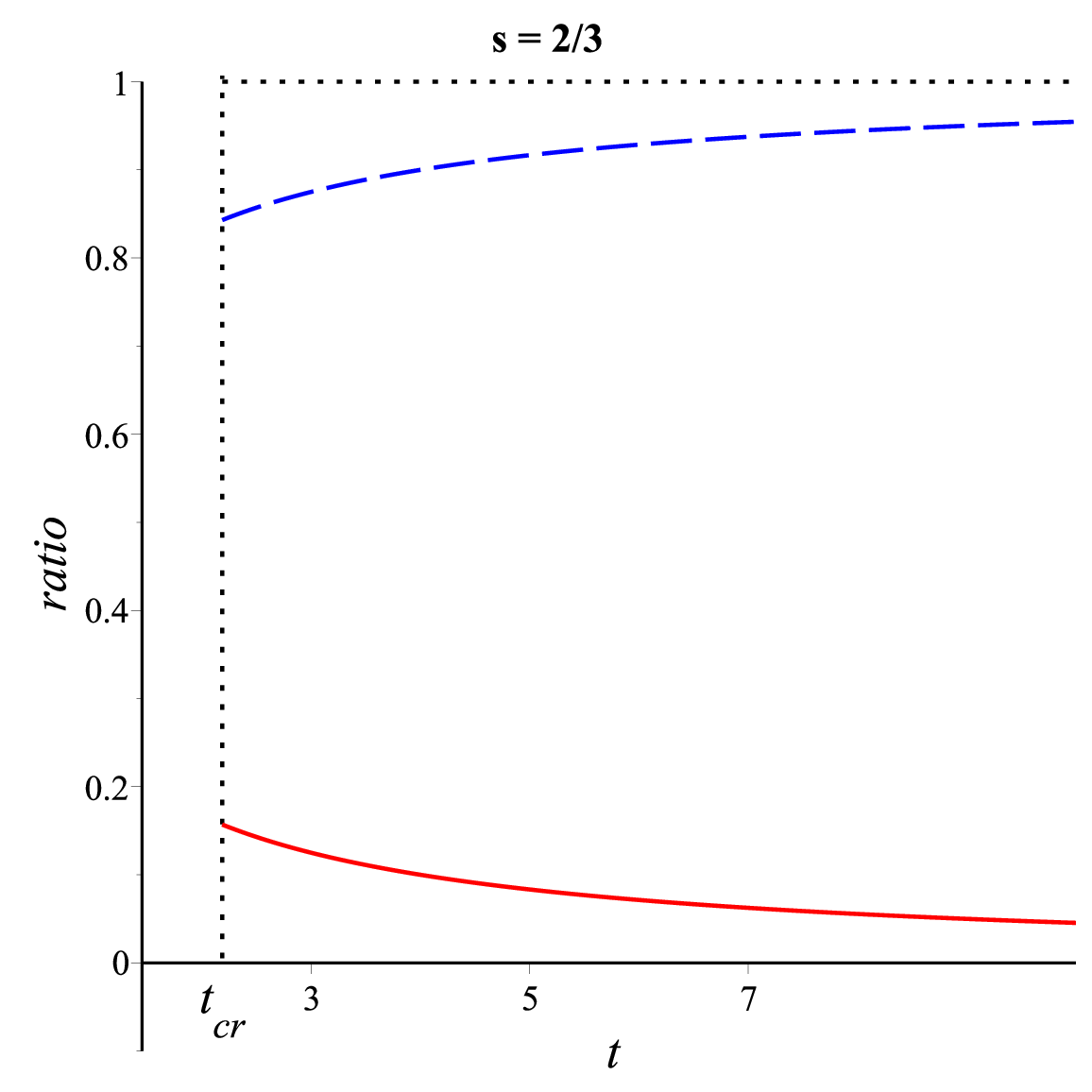}	
		\includegraphics[scale=0.21]{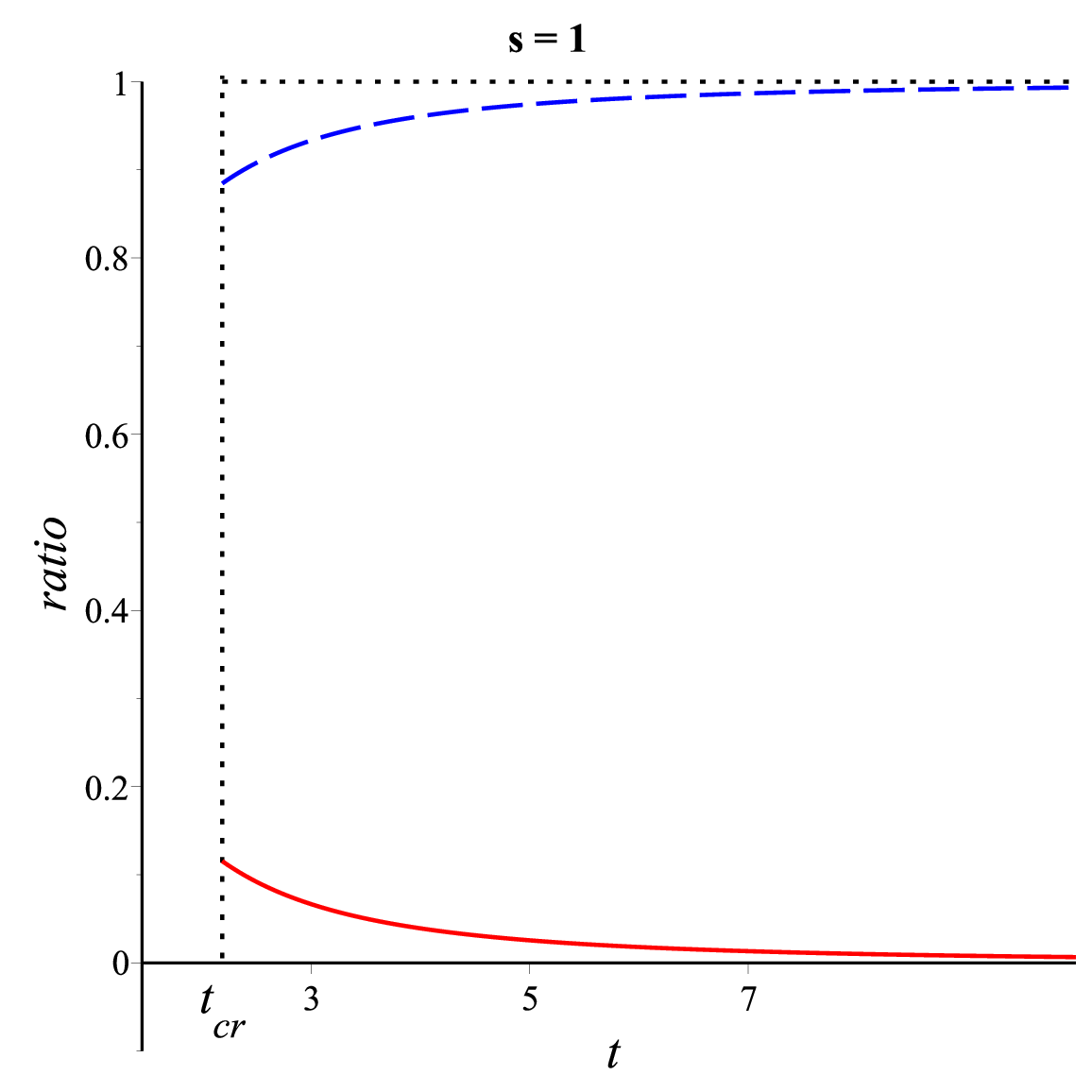}		
		\caption{The evolution of the mass functions, $M(t)$ (solid line), $\delta M(t)$ (dashed line) and $m(t)$ (dotted line) of the aforementioned models plotted in the upper panels. The profiles of the corresponding ratios, $R(t)$ (solid line) and $\delta R(t)$ (dashed line) are presented in the lower panels.}
\end{figure}} 

However, our principal objective here  is to study the masses dynamics in the physical domain for $t>t_{cr}$. Starting with the scale factors we obtain the mass functions (\ref{mat}) depending on the parameter $\kappa\in [0,\kappa_{lim}]$ where $k_{lim}$ results from Eq. (\ref{coco}).  In the case of the models with scale factors (\ref{aaa}) we set for simplicity $M_0=1$ and $t_0=1$ which means that we measure the masses in units of $M_0$ and the cosmic time in units of $t_0$. With these notations we obtain 	$\kappa_{lim}=s-\frac{1}{3}$. Then it is convenient to introduce the new parameter $k=\kappa \kappa_{lim}^{-1}$ such that
\begin{equation}
	\kappa=k\kappa_{lim}=k\left(s-\frac{1}{3}\right) \,, \quad 0<k\le 1\,,\quad s>\frac{1}{3}\,.
\end{equation}
With these preparations we obtain a simple form of the mass functions 
\begin{equation}\label{Mkm}
	M(t)=\frac{1}{t^{3s}(1-k)+tk} \,, \quad m(t)=\frac{1}{\kappa}\frac{\dot a(t)}{a(t)}=\frac{3s}{3s-1}\frac{1}{kt}\,,
\end{equation}
and $\delta M(t)=m(t)-M(t)$, understanding that this parametrization, $(s,k)$, holds for $k>0$ and $s>\frac{1}{3}$. 

{ \begin{figure}
		\centering
		\includegraphics[scale=0.21]{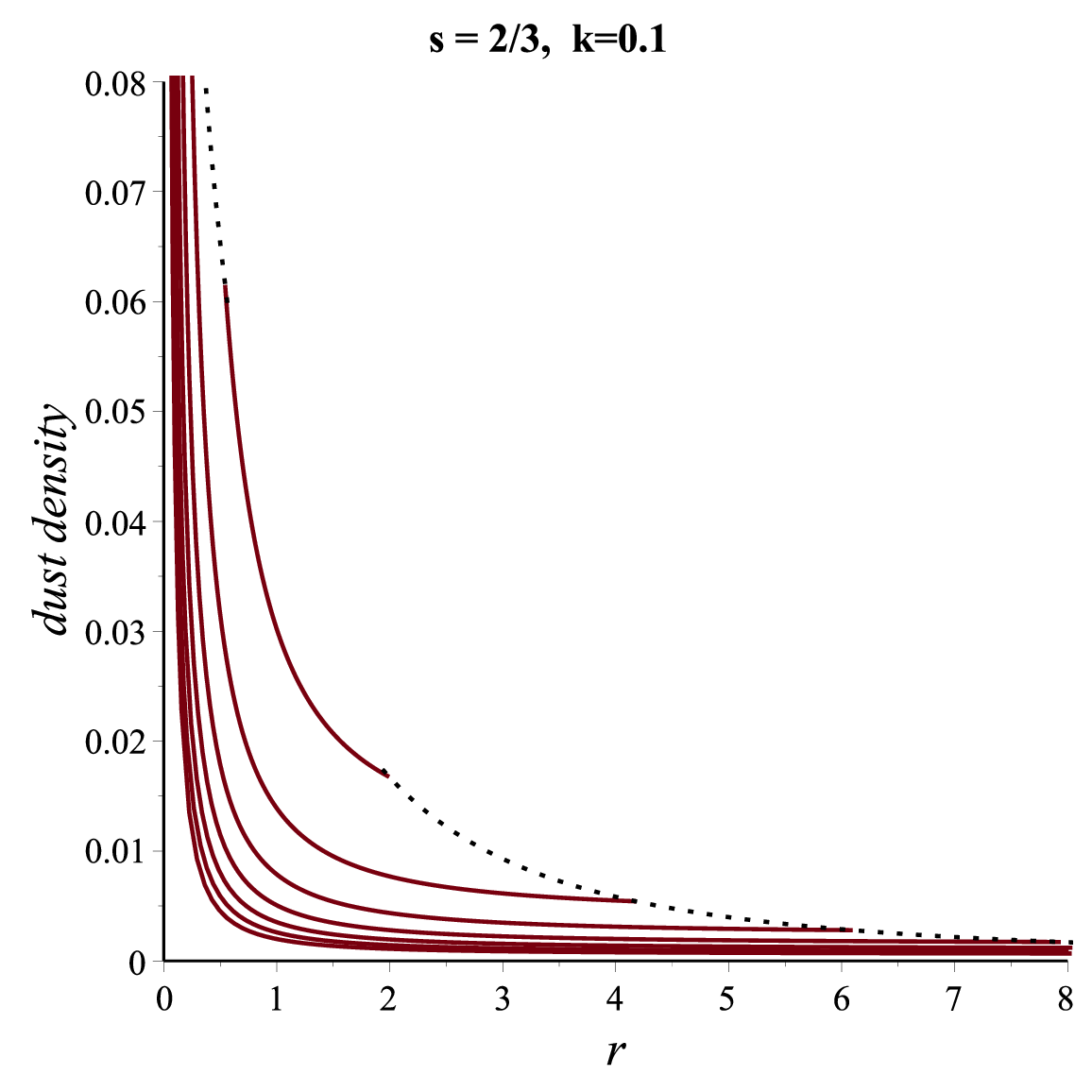}
		\includegraphics[scale=0.21]{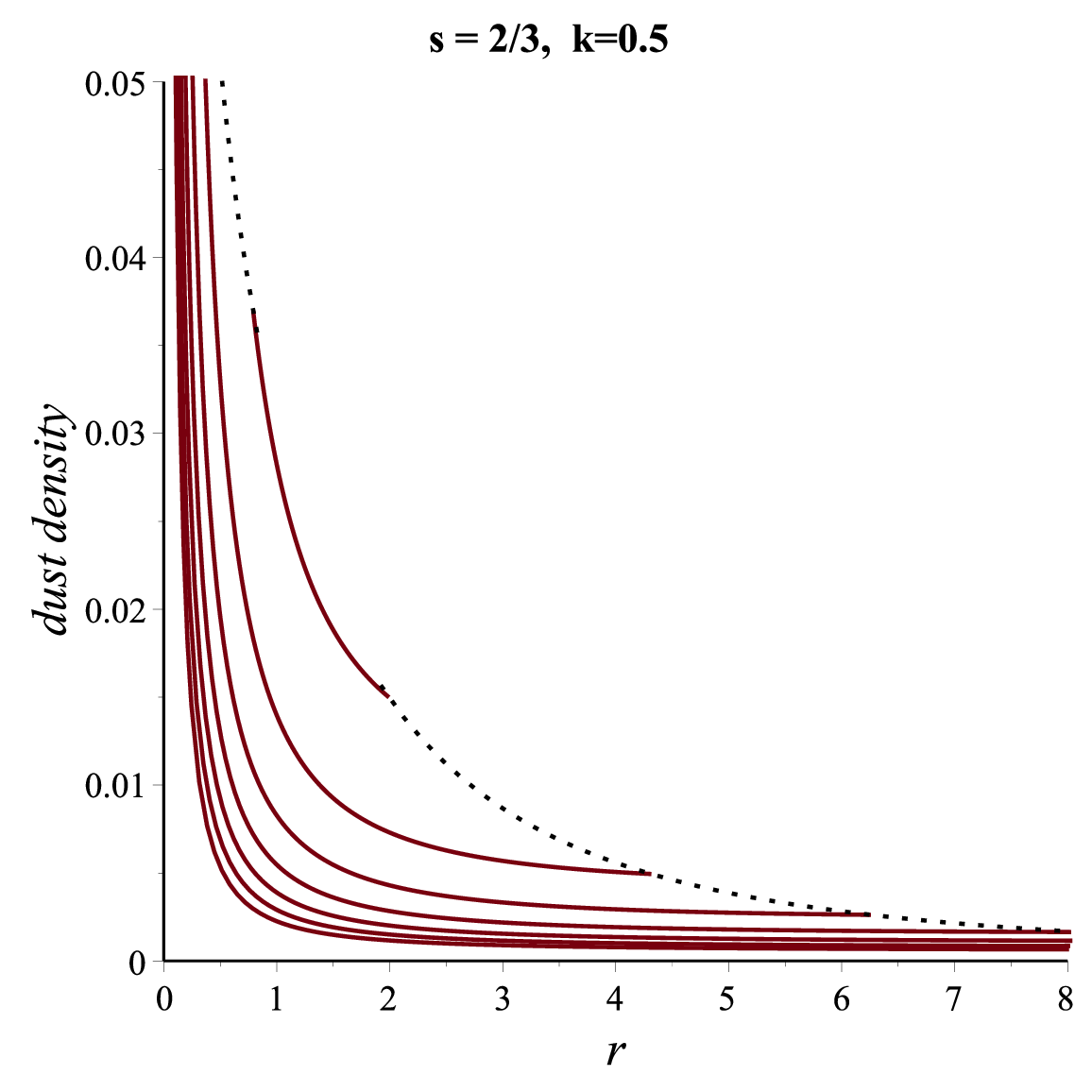}	
		\includegraphics[scale=0.21]{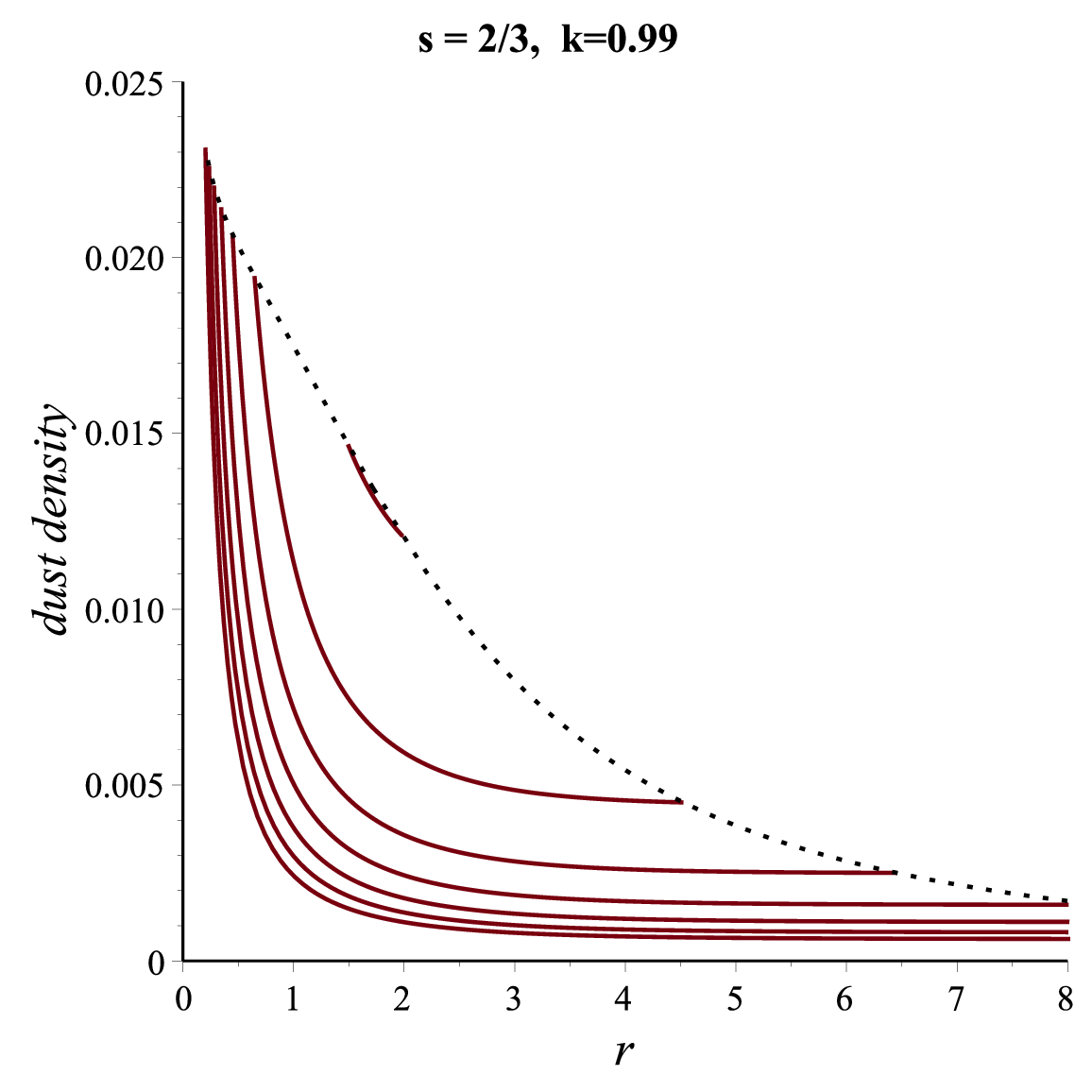}	
		\caption{Density profiles of models with parameters $(s,k)$  plotted  inside the physical domain (between the black hole and cosmological horizons) at different instants,  $t=1.01, 1, 1.5, 2, 2.5, 3$  (in units of $t_{cr}$), corresponding to the plotted profiles in descending order.  }
\end{figure}} 

Furthermore, we focus on the models of physical interest with the following asymptotic FLRW space-times,  parameters $k$ and $\kappa$ and critical times derived numerically,  
\begin{center}
\begin{tabular}{lcccc}
radiation dominated universe&$s=\frac{1}{2}$&$k=0.5$&$\kappa=0.1$&$t_{cr}=2.185$\\
matter dominated universe&$s=\frac{2}{3}$&$k=0.5$&$\kappa=0.2$&$t_{cr}=2.215$\\
Milne-type  universe&$s=1$&$k=0.5$&$\kappa=0.4$&$t_{cr}=2.169$\\
\end{tabular}	
\end{center}
 Plotting then  the mass functions of these models  as in the upper panels of Fig. 2 and the ratios 
\begin{equation}\label{Rat}
	R(t)=\frac{M(t)}{m(t)}\,, \quad \delta R(t)=\frac{\delta M(t)}{m(t)}\,,
\end{equation}
in lower panels,  we may conclude that the models with parameters $(s,k)$ describe evaporating black holes whose masses dissipate into dust such that the entire mass of the systems black hole-dust, $m(t)$, decrease monotonously, collapsing to zero for $t\to \infty$.  This means that the dust is adsorbed by the FLRW fluid which, consequently, must have a component of ordinary matter. The black hole evaporation is complete as we can prove either calculating the limits
\begin{equation}
\lim_{t\to \infty}	R(t)=0\,, \quad \lim_{t\to \infty}\delta R(t)=1\,,
\end{equation}
or plotting the ratios (\ref{Rat}) as in the lower panels of Fig. 2.

The masses of the black hole-dust system cannot be measured directly by an observer in the physical domain. The only accessible quantity is the dust density which can be measured locally.  Substituting the mass function (\ref{Mkm}) in Eq. (\ref{dust}) we obtain the dust density of the models under consideration depending on the parameters $(s,k)$. In Fig. 3 we plot successive density profiles between the black hole and cosmological horizons.  These reach their maximum values on the black hole horizon decreasing then to the cosmological one. 
We observe that when $k \to1$ then the maximal values are decreasing the dust density tending to homogeneity.

\section{Conclusions}

The general conclusion is that each $\kappa$- models with $\kappa\not=0$ describes a system formed by a black hole  surrounded by a cloud of dust hosted by the perfect fluid of the asymptotic FLRW space-time. Any such system starts to evolve  at a critical instant $t_{cr}$ when the pair of  horizons arise on the same sphere,  evolving then  creating the physical space between their spheres. Remarkably, these are genuine geometric models able to describe the black hole evaporation  without resorting to the  black hole thermodynamics. 

In what concerns the masses dynamics we conclude that in the expanding space-times under consideration the black holes evaporate integrally, dissipating their masses into dust which, in its turn, is adsorbed by the FLRW fluid. We hope that such models may be useful for   constructing realistic  models of expanding universes populated by evaporating black holes.

\end{document}